\documentclass[amsmath,amssymb,aps,prx,reprint,superscriptaddress]{revtex4-2}

\usepackage{amsmath}
\usepackage{xcolor}
\usepackage{amsfonts,amssymb}
\usepackage{graphicx}
\usepackage{hyperref}
\usepackage{academicons}
\usepackage{multirow}

\usepackage[ruled]{algorithm2e}
\usepackage{algpseudocode}

\begin{document}

\title{Deep learning of experimental electrochemistry for battery cathodes across diverse compositions}

\author{\text{Peichen Zhong}}
\email[]{zhongpc@berkeley.edu}
\affiliation{Department of Materials Science and Engineering, University of California, Berkeley, California 94720, United States}
\affiliation{Materials Sciences Division, Lawrence Berkeley National Laboratory, California 94720, United States}

\author{\text{Bowen Deng}}
\affiliation{Department of Materials Science and Engineering, University of California, Berkeley, California 94720, United States}
\affiliation{Materials Sciences Division, Lawrence Berkeley National Laboratory, California 94720, United States}

\author{\text{Tanjin He}}
\affiliation{Department of Materials Science and Engineering, University of California, Berkeley, California 94720, United States}
\affiliation{Materials Sciences Division, Lawrence Berkeley National Laboratory, California 94720, United States}

\author{\text{Zhengyan Lun}}
\affiliation{Department of Materials Science and Engineering, University of California, Berkeley, California 94720, United States}
\affiliation{School of Chemical Sciences, University of Chinese Academy of Sciences, Beijing 100049, China}

\author{\text{Gerbrand Ceder}}
\email[]{gceder@berkeley.edu}
\affiliation{Department of Materials Science and Engineering, University of California, Berkeley, California 94720, United States}
\affiliation{Materials Sciences Division, Lawrence Berkeley National Laboratory, California 94720, United States}

\date{\today}

\begin{abstract}
Artificial intelligence (AI) has emerged as a tool for discovering and optimizing novel battery materials. However, the adoption of AI in battery cathode representation and discovery is still limited due to the complexity of optimizing multiple performance properties and the scarcity of high-fidelity data. In this study, we present a machine-learning model (DRXNet) for battery informatics and demonstrate the application in the discovery and optimization of disordered rocksalt (DRX) cathode materials. We have compiled the electrochemistry data of DRX cathodes over the past five years, resulting in a dataset of more than 19,000 discharge voltage profiles on diverse chemistries spanning 14 different metal species. Learning from this extensive dataset, our DRXNet model can automatically capture critical features in the cycling curves of DRX cathodes under various conditions. Illustratively, the model gives rational predictions of the discharge capacity for diverse compositions in the Li--Mn--O--F chemical space as well as for high-entropy systems. As a universal model trained on diverse chemistries, our approach offers a data-driven solution to facilitate the rapid identification of novel cathode materials, accelerating the development of next-generation batteries for carbon neutralization.
\end{abstract}


\maketitle

\section{Introduction}

\begin{figure*}[t]
\centering
\includegraphics[width=\linewidth]{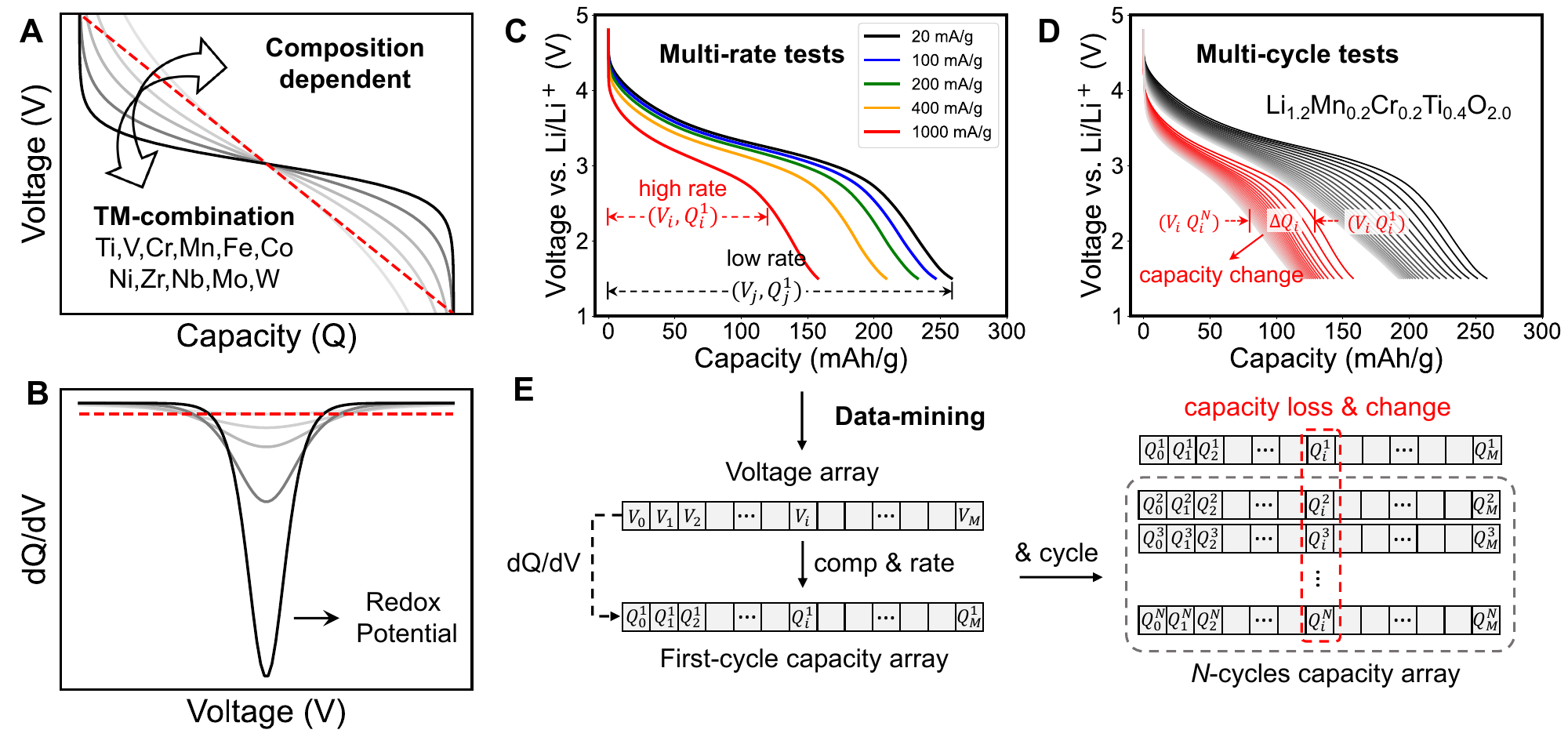}
\caption{\textbf{Discharge voltage profiles from experiments.} (A) The discharge voltage profile illustrates the relationship between capacity $Q$ and voltage $V$, which is conditioned on the composition of the cathode material. (B) The derivative quantity $dQ/dV$ is used to quantify the redox potentials of the TM. The experimental discharge voltage profiles of Li$_{1.2}$Mn$_{0.2}$Cr$_{0.2}$Ti$_{0.4}$O$_{2.0}$ DRX with (C) multi-rate tests from 20 -- 1000 mA/g and (D) multi-cycle tests from the first to the 30th cycles. (E) The parsed discharge profile is stored in a voltage array $\{V_i^N\}$, and a capacity array $\{Q_i^N\}$, where the subscript $i$ represents a point (state) on a discharge profile, and the superscript $N$ represents the cycle number.}
\label{fig:intro} 
\end{figure*}

The pursuit of carbon neutrality has become a global imperative in the face of climate change, driving the transition to renewable energy sources and the widespread adoption of electric vehicles \cite{Olivetti2017, Xie2021_carbon_neutral, Tian2021_promise}. High-performance battery cathode materials with large energy density, high-rate performance, and long cycle life are central to these advancements. The development of new cathode materials is essential to meeting the increasing demand for energy storage and advancing the electrification of transportation systems \cite{Goodenough2010}. 

Artificial intelligence (AI) has emerged as a potential tool in the discovery and optimization of novel battery materials \cite{Lv2022_ML_battery, Wang2023_nature}. By leveraging vast amounts of experimental and computational data, AI-assisted techniques can accelerate the design and synthesis of battery materials by identifying promising candidates within large chemical spaces \cite{Ahmad2018, Liow2022, chen2024accelerating}, uncovering hidden structure-property relationships via machine-learned atomistic modeling \cite{deng2023chgnet}, predicting the remaining lifespan of batteries \cite{Severson2019_natureEnergy_capacity, Jones2022_impedence_forecast, Aykol2021, hu2020battery, Sulzer2021_perspective}, and optimizing protocols for fast charge/discharge protocol \cite{Attia2020_nature_closed_loop_protocal}. These efforts significantly reduce the time and cost required for conventional trial-and-error approaches. Most recently, a battery data genome initiative has been proposed to use AI assistance to accelerate the discovery and optimization of battery materials \cite{Ward2022_battery_genome}.

Despite these advancements, current machine-learning efforts in battery research primarily focus on predicting the lifespan for a simple chemistry or within a limited chemical space, such as NMC (Ni--Mn--Co) or LFP (LiFePO$_4$). The development of exploratory machine learning for representing comprehensive compositional effects in a multi-dimensional chemical space remains underdeveloped due to the challenges associated with simultaneously optimizing multiple electrochemical properties (e.g., rate capability, cyclability, and various test voltage windows) \cite{EvanReed2022_ML_battery}. Moreover, the scarcity of high-fidelity data further hinders the progress of AI in the battery field.

Disordered rocksalt (DRX) materials have emerged as promising cathode materials that make use of earth-abundant precursors to enable scaling of Li-ion energy storage to several TWh/year production \cite{Clement2020_DRX_review}. Owing to the nearly unlimited compositional design space and considerably more complex structure-property relationship of DRX cathodes compared with conventional layered cathodes (Figure \ref{fig:intro}A), their rational design requires the extensive involvement of advanced characterization techniques (e.g., pair-distribution function analysis \cite{Key2011_PDF}, spherical-aberration-corrected transmission electron microscopy \cite{Li2021_AFM_TEM}, solid-state nuclear magnetic resonance spectroscopy \cite{Clement2018_NMR_SRO}) as well as sophisticated computational tools (e.g., high-dimensional cluster expansion and Monte Carlo simulation \cite{Zhong2022_L0L2, Barroso-Luque2022_CE_theory}). Data-driven methods offer alternative means of compositional design and optimization of materials without having to fully construct their structure-property relationships.

In light of these challenges, we developed DRXNet, an exploratory machine-learning model for the discovery and optimization of battery cathode materials. DRXNet uses composition, test current density, working voltage window, and cycle number as inputs to predict entire discharge voltage profiles. By training and testing over 19,000 experimental discharge voltage profiles of DRX materials comprising various metal species, we show that the model accurately captures the cathode electrochemistry under different test conditions. Notably, DRXNet captures accessible discharge capacity in diverse Li--Mn--O--F compositions and makes rational predictions for several high-entropy systems. As a universal model trained on diverse chemistries, DRXNet offers a data-driven solution to facilitate the rapid identification of novel cathode materials with improved energy-storage capabilities.

\begin{figure*}[t]
\centering
\includegraphics[width=\linewidth]{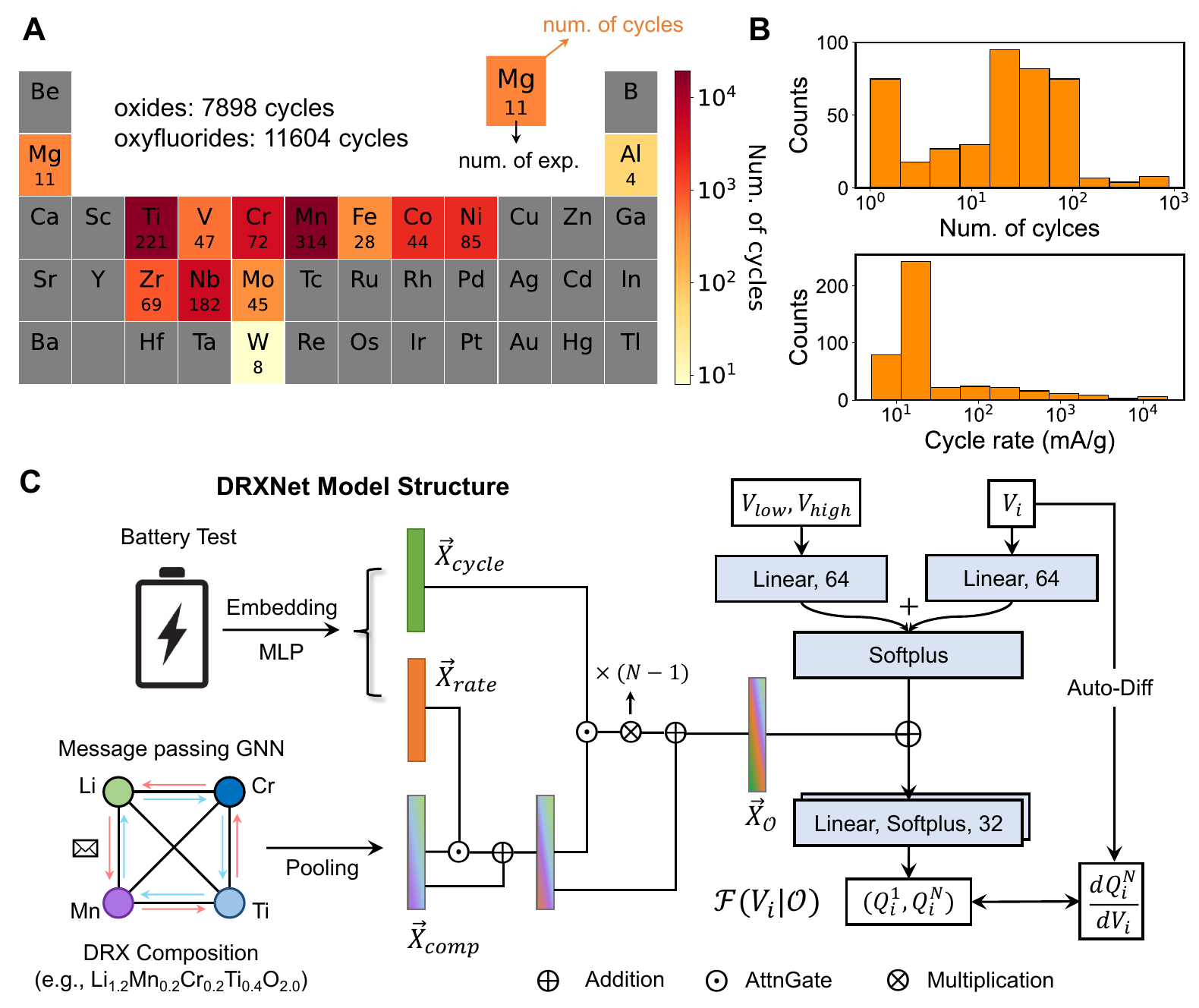}
\caption{\textbf{Description of the collected experimental dataset and model design}: 
(A) The elemental distribution of collected experimental electrochemistry data. The dataset contains 7,898 discharge profiles collected from DRX oxides and 11,604 discharge profiles from oxyfluorides. The color-coding of the boxes indicates the number of discharge profiles (cycles) on compounds that contain that specific element. The number within each elemental box represents the number of compounds with that element on which experiments were conducted. (B) A histogram of the number of cycles ($N_{\text{cycle}}$) and current density (rate) for all the individual electrochemical tests. (C)
An end-to-end pipeline that maps $Q_i = \mathcal{F}(V_{i} | \mathcal{O})$, which consists of the electrochemical condition network $\mathcal{O}$ (left) and the state prediction network $\mathcal{F}$ (right). The electrochemical condition network encodes the DRX composition, current density rate, and cycle information. The three encoded vectors are synthesized through gated-MLPs with soft attention to obtain the condition vector $\Vec{X}_{\mathcal{O}}$ \cite{Xie2018_CGCNN}. The state prediction is approximated as a forward deep neural network that takes the voltage state $V_i$ and cycling voltage window $V_{\text{low}}, V_{\text{high}}$ as inputs. The encoded condition vector $\Vec{X}_{\mathcal{O}}$ is element-wise added in the hidden layer of $\mathcal{F}$. The circled symbols are all element-wise operations. The message-passing graph neural network (GNN) is used for compositional encoding of DRX, adapted from the Roost model \cite{Goodall2020_roost}.}
\label{fig:data_model} 
\end{figure*}

\section{Results}

\subsection{Parsing discharge profiles}

Unlike conventional NMC-based layered cathodes \cite{Mizushima1980, Manthiram2020_review}, DRX materials exhibit more diverse electrochemical behavior due to the significantly larger chemical space over which they can exist and their more subtle structure involving various forms of cation short-range order \cite{Ji2019_hidden_SRO}. A prototype DRX cathode (Li$_{1+x}$M'$_a$M''$_b$O$_{2-y}$F$_y$) is composed of three primary compositional parameters: (1) the redox-active species M'; (2) the inert high-valent transition metal M'', which charge-compensates for the Li excess and stabilizes disordered structures \cite{Urban2017_PRL}; (3) fluorine, which enhances the cyclability and accommodates more Li excess without losing TM redox by reducing the anion valence \cite{Lee2017_NC}. In addition, other compositional modifications are often made to enhance capacity, rate, or cyclability. For instance, Mg doping in Mn-based oxyfluoride DRX can increase the discharge capacity while retaining a similar voltage-profile shape \cite{Zhong2020_Mg_doping}; Cr doping in Li$_{1.2}$Mn$_{0.4}$Ti$_{0.4}$O$_{2.0}$ results in comparable low-rate capacity but significantly improves the high-rate performance near the top of charge \cite{Huang2021_Cr_NatureEnergy}. These non-linear effects arising from compositional changes make both material design and machine-learning modeling challenging, thereby necessitating a comprehensive, high-fidelity dataset to address such issues. 

Figure \ref{fig:intro}A introduces the typical discharge-voltage profile in battery tests. The profile shape is tied to various factors, such as the DRX composition, applied current density rate, and degradation that may have occurred in prior cycles. Figure \ref{fig:intro}C and D show the multi-rate tests (the first cycle) and multi-cycle tests (of 20 mA/g and 1000 mA/g) of Li$_{1.2}$Mn$_{0.2}$Cr$_{0.2}$Ti$_{0.4}$O$_{2.0}$ cathode as an example. The capacity $Q$ is measured in experiments by determining the cumulative charge transferred in a galvanostatic test. Taking the derivative of $Q$ with respect to $V$, the $dQ/dV$ value can be evaluated for a given voltage profile, which is a crucial physical quantity for analyzing characteristic redox potentials from different TMs \cite{newman2021electrochemical}.

\subsection{DRX Battery Test Dataset}

We have compiled the electrochemical test data related to DRX compounds by mining electronic experimental notebooks in our research group over the past five years to construct the DRX Test Dataset (DRX-TD). The dataset contains not only results on successful materials published in several papers \cite{Lee2018_nature, Kitchaev2018_EES, Ji2019_hidden_SRO, Lun2019AEM_F_cycle, Ji2019_Ni, Lun2019_Chem, Lun2020_highentropy, Huang2021_Cr_NatureEnergy, Zhong2020_Mg_doping} but also data on less well-performing DRX compounds. This endeavor yielded a comprehensive dataset containing 19,000 discharge profiles across 16 different elements (14 metal species + O and F) from lab experiments and published literature (see \hyperref[sec:methods]{Methods}).
An individual electrochemical test is defined as a group of $N_{\text{cycle}}$ discharge profiles with a fixed current density rate, where $N_{\text{cycle}}$ is the number of cycles conducted in such a test, corresponding to the results obtained from one coin-cell. The distribution of elements in the DRX-TD is shown in Figure \ref{fig:data_model}A, where the number in each element's box represents the number of compounds with that element present for which an electrochemical test is present. The box's color indicates the total number of discharge profiles for compounds containing that element. Comprising 7,898 discharge profiles of DRX oxides and 11,604 discharge profiles of oxyfluorides, the dataset offers extensive coverage of major redox-active TMs. Figure \ref{fig:data_model}B displays histograms for the number of cycles, $N_{\text{cycle}}$, and the current rates at which experiments were performed. As is typical for exploratory research programs in a research laboratory, most of the electrochemical tests were conducted at a low current rate (20 mA/g) and for 10-100 cycles.

For each discharge profile, 100 points were uniformly sampled from the values of $V$ and $Q$, resulting in a voltage series $\boldsymbol{V} = \left[ V_1, V_2, ..., V_i, ... \right]$ and a capacity series $\boldsymbol{Q} = \left[ Q_1, Q_2, ..., Q_i, ... \right]$. The $dQ/dV$ curve was then calculated by differentiating $\boldsymbol{Q}$ with $\boldsymbol{V}$. As $dQ/dV$ is a more intrinsic property for battery materials, including this value in the modeling allows for a more representative analysis of the electrochemical performance of DRX compounds under various conditions (see \hyperref[sec:methods]{Methods}).

\subsection{DRXNet architecture}
DRXNet aims to draw a connection between chemistry and cathode performance by establishing a mapping between $\boldsymbol{V}$ and $\boldsymbol{Q}$ for arbitrary cathode compositions under various test conditions. This idea can be conceptualized as identifying a function $\mathcal{F}$ that maps cathode parameters and the voltage state $V_i$ to produce the capacity state $Q_i$ as an output. The function $\mathcal{F}$ is conditionally defined by the parameters $\mathcal{O}$, which consider the electrode composition, current rate, and cycle number
\begin{equation}
Q_i = \mathcal{F}(V_i | \mathcal{O}).
\end{equation}
We designed DRXNet with two main components, as shown in Figure \ref{fig:data_model}C:
(1) An electrochemical condition network that generates a feature vector $\Vec{X}_{\mathcal{O}}$ based on the compound's composition and  electrochemical test information; 
(2) A state prediction network to approximate the discharge state of the cathode as a function of the voltage state, $Q_i = \mathcal{F}(V_i | \mathcal{O})$, given the electrochemical conditional encoding of $\mathcal{O}$. For instance, Algorithm \ref{algo:drxnet} demonstrates how DRXNet predicts the first-cycle discharge profile of $\text{Li}_{1.2}\text{Mn}_{0.2}\text{Cr}_{0.2}\text{Ti}_{0.4}\text{O}_{2}$ at a current rate of 20 mA/g between 1.5 and 4.8 V.

\begin{algorithm}[t]
\caption{The workflow of DRXNet with an example of $\text{Li}_{1.2}\text{Mn}_{0.2}\text{Cr}_{0.2}\text{Ti}_{0.4}\text{O}_{2}$}
\label{algo:drxnet}
\begin{algorithmic}[0,width=\linewidth]
\State \textbf{Condition Inputs}:
$$
\mathcal{O} =
\begin{cases}
&\textbf{composition}  = \text{Li}_{1.2}\text{Mn}_{0.2}\text{Cr}_{0.2}\text{Ti}_{0.4}\text{O}_{2} \\
&\textbf{rate}  = 20~\text{mA/g}, \\
&\textbf{cycle} = 1
\end{cases}
$$
\State \textbf{Condition Outputs}:
$$
\Vec{X}_{\mathcal{O}_1} = \Vec{X}_{\text{comp}} + \sigma_{f_1} (\Vec{X}_{\text{comp}} ||  \Vec{X}_{\text{rate}})\cdot f_1 (\Vec{X}_{\text{comp}} ||  \Vec{X}_{\text{rate}})$$
$$
\begin{aligned}
    \Vec{X}_{\mathcal{O}_N} = \Vec{X}_{\mathcal{O}_1} + \sigma_{f_2} (\Vec{X}_{\mathcal{O}_1} ||  \Vec{X}_{\text{cycle}} ) &\cdot f_2 (\Vec{X}_{\mathcal{O}_1} ||  \Vec{X}_{\text{cycle}})\\ & \cdot \boldsymbol{W}_{n}(N-1)
\end{aligned}
$$

\Statex \hrule 
\State \textbf{Inputs}: 
$\boldsymbol{V} =  \left[1.5, ..., V_i, ..., 4.8\right] \rightarrow N$ series 
\\
\SetAlgoNoLine
\\
\For{$i = 1$ \KwTo $N$}{
\textbf{Compute} $Q_i = \mathcal{F}(V_i | \Vec{X}_{\mathcal{O}_N})$ 
}
\\
\State \textbf{Outputs}: 
$\boldsymbol{Q} =  \left[Q_1, ..., Q_i, ..., Q_N\right]$  
\end{algorithmic}
\end{algorithm}

Initially, three condition inputs (composition, rate, cycle) are encoded to represent $\mathcal{O}$. We use \texttt{Roost}, a graph neural network model proposed by \citet{Goodall2020_roost}, for compositional encoding. \texttt{Roost} takes elements as graph nodes and updates the correlation between elements through weighted message passing based on each element's fractional concentration. The nodes are initialized with elemental embedded vectors $\Vec{h}_s$ ($s$: species) from \texttt{mat2vec} to capture as much prior chemical information as possible through text mining of previously published literature \cite{Tshitoyan2019_mat2vec}. Moreover, we consider only the cation species as independent nodes in \texttt{Roost}, treating the anion-species information (fluorine) as a mean-field background, i.e., $\Vec{h}'_{\text{Li}} = \Vec{h}_{\text{Li}} + c_{\text{F}}\cdot\Vec{h}_{\text{F}}$, where $c_{\text{F}}$ is the fractional concentration of fluorine and $\Vec{h}_{\text{Li/F}}$ is the embedded vector of Li/F. Rate and cycle information is encoded using multi-layer perceptrons (MLPs).

\begin{figure*}[ht]
\centering
\includegraphics[width=0.85\linewidth]{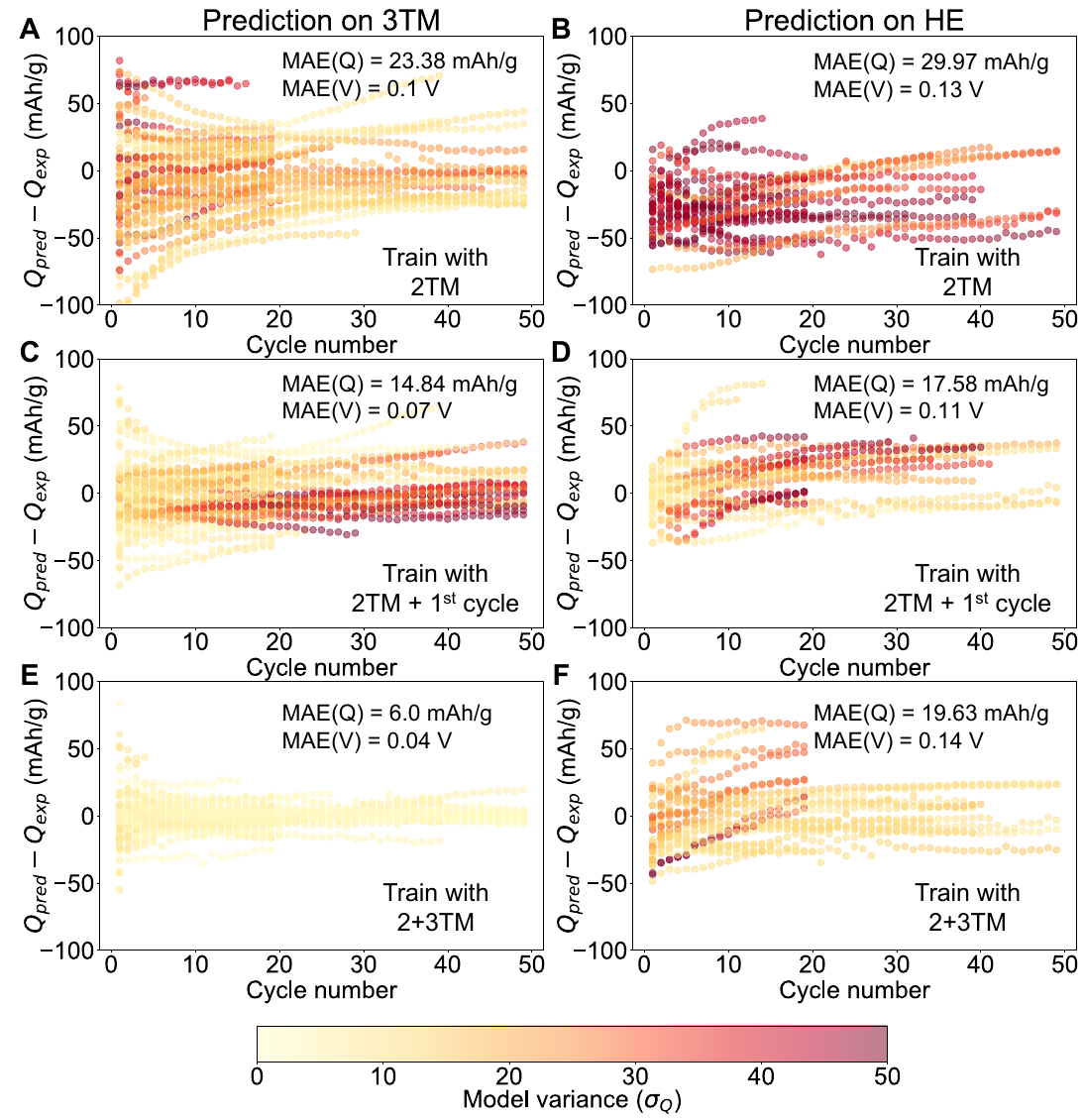}
\caption{\textbf{Error and model variance analysis of DRXNet in compositional space}: The prediction error of discharge capacity between 2.0 and 4.4 V ($y$-axis) \textit{vs.} cycle number ($x$-axis). The model variance is represented by $\sigma_Q$, a standard deviation of the ensemble of the models' prediction, which is plotted as scaled colored dots. (A)--(B): Predictions on 3TM/HE using models trained on the 2TM dataset. (C)--(D): Predictions on 3TM/HE using models trained on both the 2TM dataset and the first cycles of the 3TM/HE dataset. (E)--(F): Predictions on 3TM/HE using models trained on the 2+3TM dataset.
}
\label{fig:uncertain} 
\end{figure*}

Because the rate and cycle properties are intrinsically affected by the composition, we used gated MLPs with soft attention for electrochemical condition encoding via a hierarchical network structure \cite{Xie2018_CGCNN}. The $\Vec{X}_{\mathcal{O}_1} = \Vec{X}_{\text{comp}} + \sigma_{f_1} (\Vec{X}_{\text{comp}} ||  \Vec{X}_{\text{rate}})\cdot f_1 (\Vec{X}_{\text{comp}} ||  \Vec{X}_{\text{rate}})$ is a rate-informed feature vector, where $\sigma_f$ and $f$ represent MLPs with different activation functions and $||$ denotes the concatenation operation. In addition, the cycle-informed vector $\Vec{X}_{\mathcal{O}_N} = \Vec{X}_{\mathcal{O}_1} + \sigma_{f_2} (\Vec{X}_{\mathcal{O}_1} ||  \Vec{X}_{\text{cycle}} ) \cdot f_2 (\Vec{X}_{\mathcal{O}_1} ||  \Vec{X}_{\text{cycle}}) \cdot \boldsymbol{W}_{n}(N-1)$ is linearly dependent on the cycle number with a trainable weight $\boldsymbol{W}_{n}$. As such, the feature vector $\Vec{X}_{\mathcal{O}_1}$ is used to represent the 1st cycle and $\Vec{X}_{\mathcal{O}_N}$ is used to represent the $N$-th cycle, respectively.

Lastly, we used several MLPs to construct the state prediction network $\mathcal{F}$, as shown in Figure \ref{fig:data_model}C. $\mathcal{F}$ takes the voltage state $V_i$ and working window $V_{\text{low}}, V_{\text{high}}$ as inputs, and the $\Vec{X}_{\mathcal{O}}$ is element-wise added to the hidden layer of $\mathcal{F}$ to inform $\mathcal{F}$ of conditions $\mathcal{O}$ (see \hyperref[sec:methods]{Methods}). As such, the state prediction network $\mathcal{F}$ is constructed as a simple function mapping from the voltage state $V_i$ to the capacity $Q_i$. In addition, $(dQ/dV)_i$ is obtained by auto-differentiation of $\mathcal{F}$.

\subsection{Applicability domain}

We explore the scope of DRXNet's applicability in the realm of composition space. Determination of the applicability domain in battery machine-learning models can be challenging due to the unavailability of sufficient test data, as generating new data necessitates the synthesis of new materials or conducting battery cycling tests for weeks to months \cite{EvanReed2022_ML_battery, Sutton2020_AD}. Simply separating the sequence of voltage and capacity signals $\{V_i, Q_i\}$ into training and test sets can result in data leakage and a failure to represent the expected error in real applications. To evaluate the expressibility and generalization of DRXNet, we designed several experiments by partitioning the dataset based on compositions. The electrochemical tests with no more than two metal species (2TM, excluding Li) were designated as the training set, whereas the tests with three metal species (3TM) and higher numbers of TM components (high-entropy, HE) were assigned as test sets. For each test, an ensemble of five independent models was trained to enhance the overall prediction accuracy and robustness and to quantify the model variance. The average value is used for the prediction, and the standard deviation of the prediction from the ensemble of five DRXNet models ($\sigma_Q$) is used to represent the model variance as an approximation of how uncertain the predictions are.

A rational design of battery cathodes typically focuses on the capacity that can be delivered within a certain voltage window. Therefore, we used DRXNet to compute the voltage profiles with electrochemical test parameters in the test set and compared the delivered capacity between 2.0 -- 4.4 V of experiments and predictions within 50 cycles (see Figure \ref{fig:uncertain}). The voltage range of 2.0 – 4.4 V (vs. Li$^+$/Li) is reasonable for current electrolytes, and most commercialized cathodes such as LiFePO$_4$, LiCoO$_2$, and NMC operate within this voltage range. Our choice of this voltage range for testing model performance is aligned with these industry norms. The average voltages ($\Bar{V} 
= \sum_i V_i \Delta Q_i / \sum_i \Delta Q_i$) between 2.0 -- 4.4 V were subsequently computed. As a baseline, the mean absolute deviation (MAD) of average voltage is 0.16/0.21 V for 3TM/HE, and the MAD of discharge capacity is 36.59/38.54 mAh/g for 3TM/HE.  Figure \ref{fig:uncertain}A and B demonstrate the performance of the DRXNet models trained on the 2TM dataset and tested on the 3TM and HE datasets. Mean absolute errors (MAEs) of 0.1/0.13 V for the average voltage and 23.38/29.97 mAh/g for the capacity were obtained for the 3TM/HE test datasets, respectively. It is found that large prediction errors already occur for the first cycle and propagate into the subsequent cycles. Notably, a systematic underestimation of capacity is observed for the HE compounds (Figure \ref{fig:uncertain}B), which can be rationalized by the fact that 2TM compounds cannot capture the improved performance arising from the novel high-entropy physics \cite{Lun2020_highentropy, Zhou2023_angew}.

For practical applications, new data points can be continuously collected as experiments progress, enabling on-the-fly training with incoming data to improve predictive performance. To evaluate possible improvement with additional information specific to the system being tested, we evaluated the improvement when DRXNet is trained on a dataset containing all 2TM data and is provided with the first cycle data from 3TM/HE materials. The knowledge of just the first cycle data results in a reduction of the mean capacity error from 23.38/29.97 mAh/g to 14.84/17.58 mAh/g for 3TM/HE (Figure \ref{fig:uncertain}C and D). The enhanced performance achieved by explicitly training with the first cycle indicates that the model can better generalize cycling performance, even when experiments for a specific composition are not extensively sampled. This capability has the potential to significantly reduce the month-long timeframe typically required for electrochemical testing to identify whether a new cathode material has a desired cyclability or rate capability. Training the model with first-cycle data led to a substantial decrease in both the prediction error and model variance for the initial few cycles, although the model variance increased subsequently with the cycle number for untrained domains (Figures \ref{fig:uncertain}C and D).

To examine how data augmentation could improve the performance of DRXNet, we further trained models on the 2+3TM dataset where chemical information in addition to 2TM interactions is included. Figure \ref{fig:uncertain}E and F display the predictions on the 3TM (MAE: 6.0 mAh/g) and HE (MAE: 19.63 mAh/g) datasets. It is important to note that the models trained on 2+3TM data show an error reduction of around 10 mAh/g for the HE capacity prediction compared to the results obtained when training the 2TM model (Figures \ref{fig:uncertain}B and F), along with a significant reduction on the model variance. This finding suggests that the 2TM dataset is inadequate for extracting relevant information and generalizing it to other compositions. The scaling to electrode material with a high number of components necessitates capturing more than 2TM correlations or interactions in training the graph neural network. Failure to do so may lead to systematic prediction errors, as demonstrated in Figure \ref{fig:uncertain}B. When the model is able to acquire sufficient chemical domain knowledge  (e.g., 2+3TM-model), it becomes feasible to extrapolate the electrochemical properties of high-component electrodes, which is evidenced in Figure \ref{fig:uncertain}F with reduced prediction error as well as model variance, and only a few outlier experiments exhibit large errors.

\begin{figure*}[t]
\centering
\includegraphics[width=\linewidth]{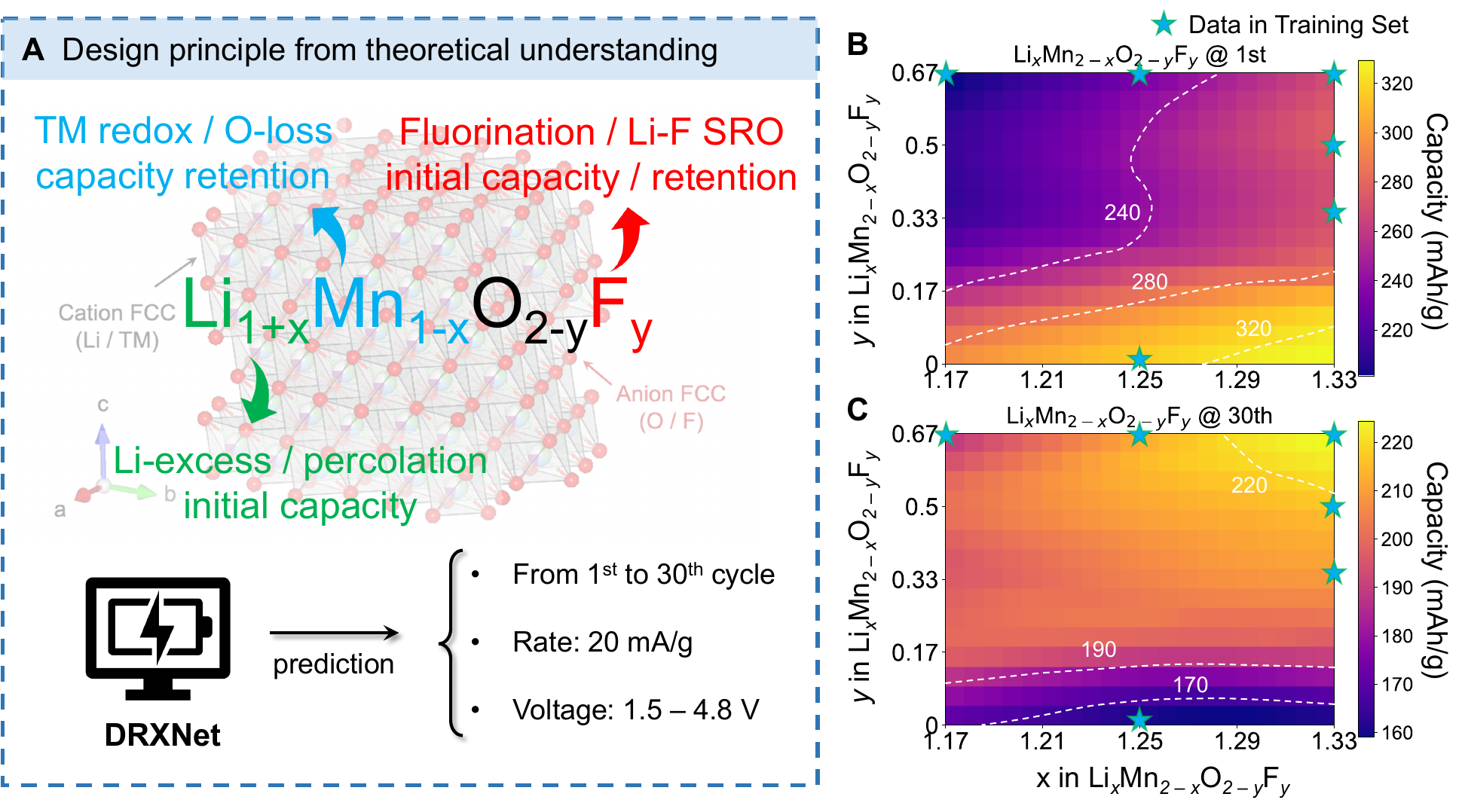}
\caption{\textbf{Illustration of predictions of discharge capacity in Li--Mn--O--F DRX systems}: (A) Compositional design principle includes the optimization of Li-excess content, TM-redox and Li-F short-range order (SRO) \cite{Lun2019_Chem}.} (B, C) Prediction of discharge capacity in the Li--Mn--O--F chemical space for the 1st (B) and 30th (C) cycle between 1.5 -- 4.8 V at a current density rate of 20 mA/g. The blue stars indicate the compositions included in the training set.
\label{fig:LMOF} 
\end{figure*}

\subsection{Recover design principles in Li--Mn--O--F chemical space}

We present several examples to illustrate how DRXNet learns the underlying cathode chemistry and assists in designing new materials, where the models used for these applications are pretrained on all discharge profiles. As an attractive earth-abundant, non-precious TM, Mn is of considerable interest for designing next-generation cathode materials \cite{Lee2018_nature}. \citet{Lun2019_Chem} proposed three primary design degrees of freedom for Mn-based DRX (Figure \ref{fig:LMOF}A): (1) the Li-excess content, which controls the presence of a percolating network facilitating Li diffusion; (2) the Mn content, as achieving high capacity with a low amount of Mn requires a large amount of oxygen redox, leading to poor cyclability; and (3) the fluorine content, which lowers the total cation valence and provides greater freedom to optimize the Li and Mn content. Fluorine modifies cation short-range order (SRO) through the strong Li--F attraction and lowers the initial capacity \cite{Ouyang2020_F_perco, Zhong2020_Mg_doping}, but can increase stability at high voltage charging \cite{Li2021_AFM_TEM}. These principles are highly correlated and exert non-linear effects on performance.

We used DRXNet to predict the discharge capacity of DRX compounds between 1.5 and 4.8 V at a current rate of 20 mA/g for the 1st and 30th cycles. The results, as a function of Li and F content, are shown in Figures \ref{fig:LMOF}B and C. The Mn content and valence follow directly from the Li and F content. The effect of fluorine on performance, extensively characterized experimentally, is well captured by our model: A higher F content ($y$ in O$_{2-y}$F$_{y}$) results in a lower discharge capacity for the 1st cycle but a higher capacity for the 30th cycle, consistent with its documented role in promoting surface stability \cite{Li2021_AFM_TEM}. In particular, cation-disordered Li$_{1.333}$Mn$_{0.667}$O$_2$ (bottom right corner of Figure \ref{fig:LMOF}C) is predicted to have the highest capacity ($>320$ mAh/g) for the first cycle but the lowest capacity for the 30th cycle. In this compound, the valence of Mn is 4+, and all capacity originates from oxygen. Such a large amount of O-redox leads to rapid capacity fade consistent with the experimental observations on disordered Li$_2$MnO$_3$ reported in Ref. \cite{Kataoka2018_dis_Li2MnO3}.

To provide some context for the extrapolation capability of DRXNet, we have illustrated the compositions in the training dataset with blue stars in Figures \ref{fig:LMOF}B and C. From this, it can be observed that even with a limited distribution of training points on the composition map, DRXNet offers reasonably consistent predictions that seem to be in line with the experimental observations beyond the training points. As DRXNet is trained on various compositions beyond the Li--Mn--O--F chemical space, the ability to extrapolate to other domains can be attributed to the transfer learning from other F- and non-F-containing compounds. The example in this section demonstrates how practitioners can generalize the design principles from a data-driven perspective purely starting from the data mined from experiments.

\subsection{Exploratory search for high-entropy cathodes}
\begin{figure*}[ht]
\centering
\includegraphics[width=\linewidth]{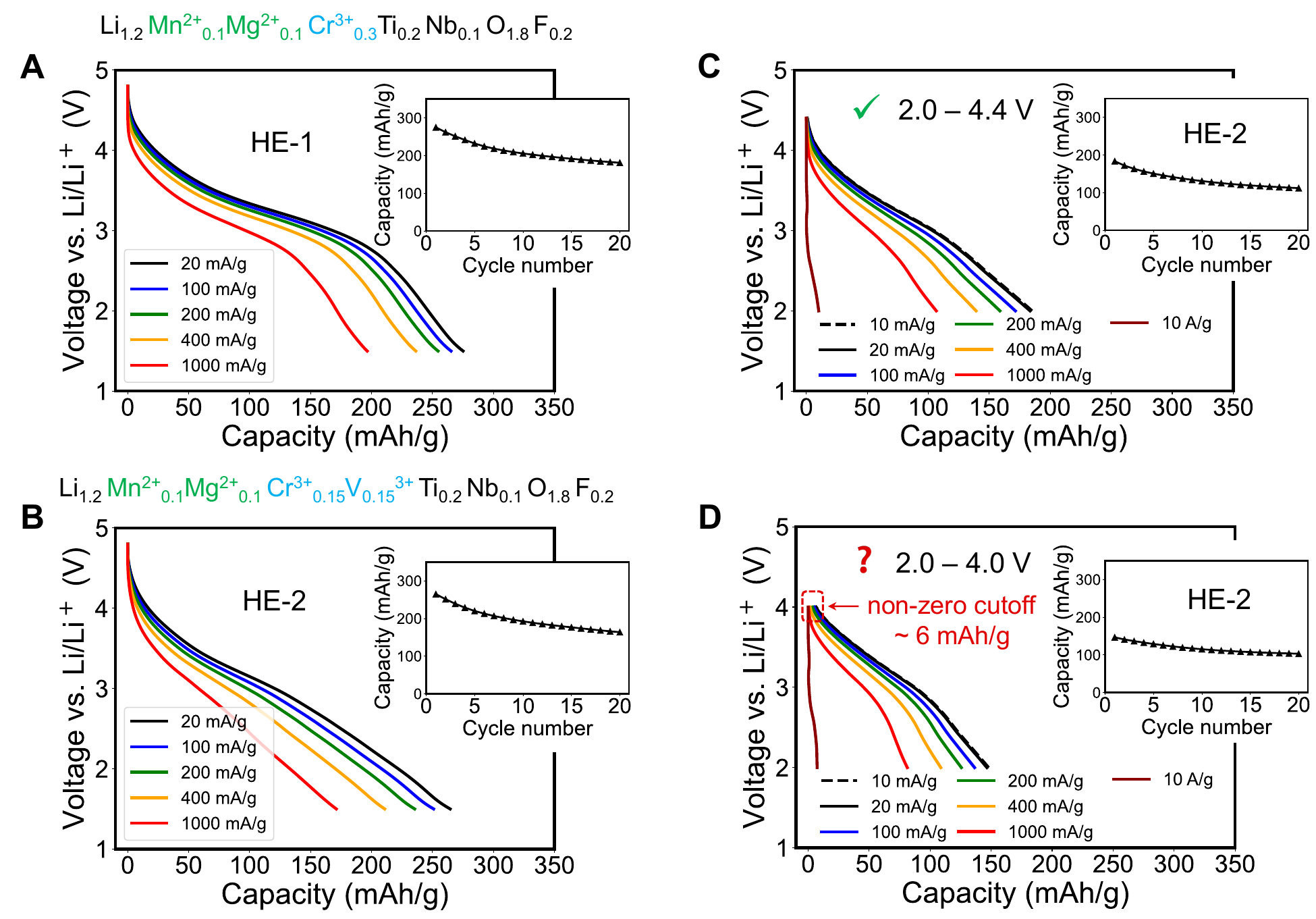}
\caption{Predicted discharge voltage profiles of two high-entropy DRX materials. (A) Li$_{1.2}$Mn$_{0.1}$Mg$_{0.1}$Cr$_{0.3}$Ti$_{0.2}$Nb$_{0.1}$O$_{1.8}$F$_{0.2}$ (HE-1) and (B) Li$_{1.2}$Mn$_{0.1}$Mg$_{0.1}$Cr$_{0.15}$V$_{0.15}$Ti$_{0.2}$Nb$_{0.1}$O$_{1.8}$F$_{0.2}$ (HE-2) with various current densities (from 20 mA/g to 1000 mA/g) between voltage window of 1.5 --4.8 V. The inset displays the cycled discharge capacity at a current density of 20 mA/g. HE-2 with various current densities (from 10 mA/g to 10 A/g) between voltage windows of (C) 2.0 --4.4 V and (D) 2.0 -- 4.0 V. }
\label{fig:HE12} 
\end{figure*}

High-entropy DRXs are composed of many species and present a vast chemical space to explore for battery materials discovery. We used DRXNet for virtual high-throughput screening considering redox-compatible species from the bivalent (Mn$^{2+}$, Fe$^{2+}$, Ni$^{2+}$, Mg$^{2+}$) and trivalent group (Mn$^{3+}$, Cr$^{3+}$, V$^{3+}$, Fe$^{3+}$). Two case studies of predicted high-entropy DRXs are presented: Li$_{1.2}$Mn$_{0.1}$Mg$_{0.1}$Cr$_{0.3}$Ti$_{0.2}$Nb$_{0.1}$O$_{1.8}$F$_{0.2}$ (HE-1) and Li$_{1.2}$Mn$_{0.1}$Mg$_{0.1}$Cr$_{0.15}$V$_{0.15}$Ti$_{0.2}$Nb$_{0.1}$O$_{1.8}$F$_{0.2}$ (HE-2). The discharge profiles predicted with DRXNet under various current densities are shown in Figures \ref{fig:HE12}A and B. A more comprehensive collection of predictions for other compositions is included in Figure \hyperref[sec:SI]{S6}.

For HE-1, DRXNet predicts a discharge capacity of 276 mAh/g at a current rate of 20 mA/g. The compound delivers its largest discharge capacity near 3V and transitions to a higher voltage slope below 3V, a phenomenon that has been widely observed in Mn redox and/or Cr redox-based DRXs \cite{Ren2015, Huang2021_Cr_NatureEnergy, Lun2020_highentropy}. HE-1 is predicted to have an unusually high rate capability for a DRX compound when discharging. A capacity of 196 mAh/g is estimated at 1000 mA/g, which is  $71\%$ of the capacity at 20 mA/g. Previous work has demonstrated that multi-elemental substitution (i.e., high-entropy strategy) frustrates the unfavorable short-range order that leads to poor Li kinetics. In addition, the incorporation of Cr and its migration as Cr$^{6+}$ at high voltage creates a more extended 0-TM network for Li transport. Both of these features improve the Li diffusion kinetics \cite{Huang2021_Cr_NatureEnergy, Lun2020_highentropy}.  DRXNet clearly learns those benefits and extrapolates rationally into electrochemistry prediction of the high-entropy compositions.

As a comparison to HE-1, we formulated HE-2 with partial V$^{3+}$ to Cr$^{3+}$ substitution. The change in the shape of the voltage profile due to the low potential of V$^{5+}$/V$^{3+}$ reduction is well captured by DRXNet as shown in Figure \ref{fig:HE12}B and $dQ/dV$ curves in SI. It is clearly demonstrated that with V$^{3+}$ incorporation, a nearly constant slope can be observed down to the low-voltage region, which is characteristic for reported V-based DRX cathodes \cite{Nakajima2017_LVN, chen2015_LVO}. Nevertheless, similar to Cr$^{6+}$, V$^{5+}$ can migrate into the tetrahedral sites to enhance Li transport, which benefits the rate capability \cite{Nakajima2017_LVN}. Consistently, with this concept, HE-2 is predicted to retrain 171 mAh/g capacity at 1000 mA/g ($64\%$ of the 266 mAh/g capacity at 20 mA/g), which is superior to the majority of the DRX cathodes reported to date.

The inset plots in Figures \ref{fig:HE12}A and B show the predicted discharge capacity of HE-1 and HE-2 for 20 cycles. The capacity drop in the first five cycles is predicted to slow down upon further cycling. This result is in full agreement with experimental findings, which indicate some of the irreversibility in the initial cycles, such as cathode--electrolyte interface formation \cite{Edstrom2004_SEI}. These examples illustrate how practitioners can effectively use DRXNet to navigate the extensive chemical space of high-entropy DRXs and identify promising candidates for cathode design and optimization.

\subsection{Electrochemical conditions}

We further tested the depth and transferability of DRXNet's predictive capabilities by varying the HE-2 discharge voltage window and cycling rate, which are typical parameters varied in the investigation of a new cathode material. Figure \ref{fig:HE12}C displays the discharge profiles between 2.0 -- 4.4 V, with two additional rates tested (10 mA/g for a low rate and 10$^4$ mA/g for an extremely high rate). These conditions are infrequently incorporated into our training data. The 10 mA/g exhibits a discharge profile very similar to that obtained at 20 mA/g, which is entirely consistent with typical experimental findings, as the discharge process at such a low rate exhibits a reduced overpotential and is closer to the equilibrium. The 10 A/g rate discharge profile demonstrates a sharp drop in voltage, reasonably indicating poor performance at this extremely high rate. Some unphysical predictions start to appear when the model is tested to predict the discharge profiles between 2.0 -- 4.0 V. As Figure \ref{fig:HE12}D shows a small non-zero offset $\sim 6$ mAh/g for the  20 mA/g rate profile appears at the onset of discharge (4.0 V). Since the start of the capacity curve at the upper level of the voltage cutoff is not formally enforced to zero by the model, but emerges from the linear embeddings of the voltage state $V_i$ with the voltage window $[V_{\text{low}}, V_{\text{high}}]$, an offset can be created when there is not enough data for that specific voltage window.

Based on the tests, our primary conclusion is that DRXNet exhibits a reasonable ability to learn the cathode material's chemical information in the latent space and generalize to test conditions that are included in the dataset. However, for test conditions that the model has not or rarely encountered (e.g., experiments with $V_{\text{high}} < 4.0$ V), discrepancies or unphysical profiles may still arise. This highlights the data scarcity issue, which is typical for human-generated experimental conditions, which are biased toward what is needed to demonstrate performance rather than what is optimal for model training \cite{Raccuglia2016_fail_exp}.

\section{Discussion}

Most machine learning approaches predicting battery performance have been focused on predictions for a specific chemistry or limited chemical space of commercialized cathodes, typically, the remaining useful life forecasted from the initial cycles \cite{Severson2019_natureEnergy_capacity, Zhu2022, Roman2021_NMI}. However, the nature of battery cathode material discovery and optimization lies in a broad domain of chemistries, which is more challenging for AI as it needs to capture the direct (e.g., voltage) and indirect effect (e.g., cycle life) of chemical changes \cite{EvanReed2022_ML_battery}. Recent studies have demonstrated the feasibility of building universal models for atomistic modeling by harnessing more than ten years of \textit{ab-initio} calculations spanning the periodic table \cite{Jain2013, Chen2022_m3gnet, deng2023chgnet, Merchant2023_gnome, batatia2023foundation}. It becomes a logical extension to envision universal models for the experimental discovery of battery materials by leveraging the wealth of both \textit{ab-initio} calculation and experimental data generated on cathode materials worldwide \cite{Ward2022_battery_genome, Hong2021_reduce}. In this work, we propose an end-to-end training pipeline to encode and learn the (electro)chemical information of cathode materials from voltage profiles. Focused on DRX cathodes, we data-mined years of lab-generated experimental discharge voltage profiles and trained a universal machine-learning model (DRXNet) to make predictions across diverse compositions. This was achieved through a novel model design consisting of an electrochemical condition network $\mathcal{O}$ and a state prediction network $\mathcal{F}$.

The design of the two networks promotes modularity in the architecture, streamlining the optimization and interpretation of each network individually and their learned features. For instance, composition is an intrinsic property of the synthesized cathode materials, and the encoding of such features is independent of other factors such as current density and cycle status, rationalizing our approach to first extract the composition-only feature $\Vec{X}_{\text{comp}}$ via a GNN. Although it remains a challenge that the composition may change as a function of current density and cycle status due to TM dissolution and the irreversible reaction of lithium outside the cathode, DRXNet encompasses these factors into the rate- and cycle-informed feature vector representations. By leveraging a ResNet-inspired architecture using skip connections \cite{He2016_resnet}, we achieve a more effective synthesis of the feature vector within the latent space. This design allows for a direct connection between the rate-informed feature, $\Vec{X}_{\mathcal{O}_1}$, and the prediction of the first cycle capacity. Such architecture has been proven to boost model training and alleviate the well-known gradient vanishing issues. 

Given the inherent sequential nature of battery testing data -- where possessing information from the $N$-th cycle implies the availability of data from the first cycle -- it becomes crucial to design features that reflect this causality. This insight leads to the formulation of the cycle-informed feature, $\Vec{X}_{\mathcal{O}_N}$. This feature accentuates the difference between the first and the $N$-th cycles, guiding the prediction for the $N$-th cycle capacity, as detailed in Eq.\eqref{eq:feature_vectors}.
Consequently, our loss function is constructed for multi-task learning with both terms for the first and $N$-th cycle capacities, ensuring the causal relationships in cycle-dependent capacity predictions (refer to Eq.\eqref{eq:loss_function}). Through an ablation study on whether to include the first cycle term, $\ell(Q^1)$, in the loss function or not, we found that the model without $\ell(Q^1)$ tends to be underfitted (more details in Figure  \hyperref[sec:SI]{S8}). Our incorporation of loss terms for both the first and $N$-th cycle capacities enhances the model expressibility, which is a crucial factor in the optimization of battery materials.

In addition, the modular design of the electrochemical condition network $(\mathcal{O})$ provides flexibility for the feature representation when expanding the model to include other information. The training dataset, being derived from our own experimental results, does not encompass testing parameters such as particle size, electrolyte type, synthesis variations, etc. Since the battery electrodes were fabricated in our laboratory using standardized recipes and methodologies, these factors have been coarsely integrated into the compositional model and are treated as constants across our dataset. Currently, the model does not include features to capture structural information (crystal structure, short-range order, etc.).  In DRX compounds, short-range order is known to influence performance and to the extent that this is not a direct consequence of composition, but modified by synthesis parameters its effects are not accounted for \cite{Cai2022_thermal}. In principle, researchers can choose to include such factors to design the electrochemical feature vector, depending on the specific problem they are addressing. Given the vast amount and complexity of these properties, a synthetic data collection approach is necessary. Data-mining techniques, such as text mining and figure mining, can automatically retrieve valuable experimental information from decades of published literature \cite{Kononova2021_textmining_review, Baibakova2022_figure_extraction}, though a challenge with aggregating diverse data from literature is the numerous hidden and unspecified variables relevant to materials synthesis and electrochemical testing. Looking forward, automated labs can address both data scarcity and transparency issues by enabling more extensive exploration of the experimental space and even better collect data from "failed" experiments \cite{Nathan2021_ALab_review, Stach2021, Szymanski2023, Lunt2024}.

In conclusion, DRXNet represents a step forward in developing machine-learning models for battery materials research. By continuously refining the model and incorporating additional data and parameters, we anticipate that such a machine-learning framework will play an increasingly critical role in discovering and optimizing next-generation battery materials.

\section{Methods}
\label{sec:methods}
\subsection{Data collection}
We collected coin-cell electrochemical test data from our lab starting in 2016 and converted them into a digital format (\texttt{.json}). Each \texttt{.json} file contains information on one individual electrochemical test, including the electrode composition, electrode mass (g), active mass (g), test current rate (mA/g), low and high voltage value of the working window (V), and charge/discharge profiles of $N_{\text{cycle}}$ collected cycles. The compositions used for the model training were taken as the targeted composition in experiments. For the fraction of our data set which was previously published, the composition values were typically confirmed by Inductively Coupled Plasma (ICP) analysis. For these compounds, the feature vectors of the targeted compositions and ICP-analyzed composition exhibit $\geq 99.7\%$ in cosine similarity as shown in \hyperref[sec:SI]{Supplementary Information}, which supports using the targeted composition for the general prediction purpose. Nonetheless, minor variations between the actual composition and the target composition can be a source of noise in the data. 

For the in-house battery tests, the CR2032 coin cells were assembled using commercial 1 M LiPF$_6$ in an ethylene carbonate and dimethyl carbonate solution (volume ratio 1:1) as the electrolyte, glass microfiber filters (Whatman) as separators, and Li-metal foil (FMC) as the anode. The coin cells were tested on an Arbin battery cycler at room temperature. The cathode consisted of a mixture of active material (DRX), Super C65 carbon black, and polytetrafluoroethylene (PTFE). The capacity signal, collected in units of Ah from the Arbin battery cycler, was normalized to mAh/g using the mass of the active material (active mass). The data from the failed tests (e.g., Arbin cycler breakdown, electrolyte failure, strong signal fluctuations, etc.) were removed from the dataset (see Figure \hyperref[sec:SI]{S1} for examples).

To enhance the generalization and expressibility of DRXNet, we expanded the dataset by figure mining published voltage profiles in related systems not covered by our lab tests (see Table  \hyperref[sec:SI]{S1}), which was accomplished using the \texttt{WebPlotDigitizer} \cite{Rohatgi2022}. We used the \texttt{UnivariateSpline} method to denoise and resample the experimental profiles and compute the $dQ/dV$ curves. One hundred points were uniformly sampled to form the voltage series $\boldsymbol{V} = \left[ V_0, V_1, ..., V_i, ... \right]$ for each discharge profile, and the capacity series and $dQ/dV$ series were calculated accordingly from $\boldsymbol{V}$. 

\subsection{Model design}

\subsubsection{Preliminaries}
A linear layer  with trainable weight $\boldsymbol{W}$ and bias $\boldsymbol{b}$ is defined as 
\begin{equation}
  L(\Vec{X}) = \Vec{X}\boldsymbol{W} +\boldsymbol{b}.
\end{equation}
For simplicity of notion, each $L$ represents different trainable weights in the following equations.

\subsubsection{Compositional encoding}

For elemental information, each element is first embedded into a 200-dimensional vector using \texttt{mat2vec} \cite{Tshitoyan2019_mat2vec}. The \texttt{Roost} (Representation Learning from Stoichiometry) model is used for compositional encoding \cite{Goodall2020_roost}, which is a  graph neural network (GNN) with message passings as follows:
\begin{equation}
\begin{aligned}
    \Vec{h}_i^{t+1} &= \Vec{h}_i^{t} + \sum_{j,m} a_{i,j}^{t,m} \cdot \sigma_g \circ L_c \left(\Vec{h}_i^{t} ||  \Vec{h}_j^{t} \right) ,\\
    a_{i,j}^{t,m} &= \frac{w_j \exp(e_{i,j}^{t,m})}{\sum_k w_k \exp(e_{i,k}^{t,m})}, ~e_{i,k}^{t,m} = \sigma_g \circ  L_a\left(\Vec{h}_i^{t} ||  \Vec{h}_j^{t} \right).
\end{aligned}
\end{equation}
In these equations, $\Vec{h}_i^{t}$ represents the $t$-th hidden layer for the $i$-th element; $||$ denotes the concatenation operation; and the soft-attention coefficient $a_{i,j}^{t,m}$ describes the interaction between elements $i$ and $j$, with $m$ as the index of multi-head attention. $L_c$ and $L_a$ denote the linear layer for the core and attention layer, respectively. The fractional concentration $w_j$ of element $j$ depends on the specific compound (e.g., $w_j= 0.6/0.1/0.1/0.2$ for Li/Mn/Cr/Ti in Li$_{1.2}$Mn$_{0.2}$Cr$_{0.2}$Ti$_{0.4}$O$_{2.0}$). $\sigma_g$ is the \texttt{SiLu} activation function. After $n$ graph convolution layers, the encoded composition vector $\Vec{X}_{\text{comp}}$ is obtained by average pooling over the elements with weighted attention 
\begin{equation}
    \Vec{X}_{\text{comp}} = \text{Pooling}\left[ \frac{w_i \exp\left(\sigma_g \circ L_a(\Vec{h}_i^{n})\right)}{\sum_k  \exp\left(\sigma_g \circ L_a(\Vec{h}_i^{n})\right)} \cdot \left(\sigma_g \circ L_c(\Vec{h}_i^{n})\right) \right]
\end{equation}

\subsubsection{Electrochemical condition encoding}
The electrochemical test primarily involves two pieces of information: the current density rate and cycle number. We use MLPs to encode the rate and cycle number:
\begin{equation}
    \Vec{X}_{\text{rate}} = \sigma_g \circ L(\text{rate}), ~\Vec{X}_{\text{cycle}} = \sigma_g \circ L(\text{cycle}).
\end{equation}
As the actual rate and cycle performance are strongly correlated with cathode materials, the relationship between the composition, rate, and cycle is synthesized using gated-MLPs with soft attention\cite{Xie2018_CGCNN}:
\begin{equation}
\begin{aligned}
    \Vec{X}_{\mathcal{O}_1} &= \Vec{X}_{\text{comp}} + \sigma_{f_1} (\Vec{X}_{\text{comp}} ||  \Vec{X}_{\text{rate}}) \cdot f_1 (\Vec{X}_{\text{comp}} ||  \Vec{X}_{\text{rate}}) \\
    \Vec{X}_{\mathcal{O}_N} &= \Vec{X}_{\mathcal{O}_1} + \sigma_{f_2} (\Vec{X}_{\mathcal{O}_1} ||  \Vec{X}_{\text{cycle}} ) \cdot f_2 (\Vec{X}_{\mathcal{O}_1} ||  \Vec{X}_{\text{cycle}}) \\
    & \quad \cdot \boldsymbol{W}_{n}(N-1)
    \label{eq:feature_vectors}
\end{aligned}
\end{equation}
where $\sigma_f = \sigma_s \circ B \circ L$ is an MLP, $\sigma_s$ is the \texttt{Sigmoid} activation function, and $f = \sigma_g \circ B \circ L$ is an MLP with \texttt{SiLu} activation function $\sigma_g$. The \texttt{BatchNormalization} layer $B$ is added before the activation function.  In this equation, $\Vec{X}_{\mathcal{O}_1}$ is a feature vector jointly determined by the composition and rate information, which is used to predict the first cycle property. $\Vec{X}_{\mathcal{O}_N}$ is a feature vector jointly determined by the composition, rate, and cycle information, which is used to predict the $N$-th cycle property. The difference between $\Vec{X}_{\mathcal{O}_1}$ and $\Vec{X}_{\mathcal{O}_N}$ is linearly dependent on the number of cycles with a trainable weight $\boldsymbol{W}_{n}$, allowing the model to learn cycle performance contrastively.

\subsubsection{State prediction network}
The state prediction network ($\mathcal{F}$) takes the inputs of voltage state ($V_i$) and outputs the discharge-capacity state ($Q_i$) 
\begin{equation}
    Q_i = \mathcal{F}\left(V_i | \mathcal{O}\right) .
\end{equation}
In practice, the voltage profile is measured within the applied voltage window [$V_{\text{low}}, V_{\text{high}}$]. To accommodate the voltage window in the discharge state prediction, the first layer in $\mathcal{F}$ is encoded via an MLP:
\begin{equation}
\begin{aligned}
    \Vec{Z}_i^0 & =  L \circ \sigma_{\mathcal{F}} \circ \left[  L(V_{\text{low}}, V_{\text{high}}) + L(V_i)  \right],
\end{aligned}
\end{equation}
where $\sigma_{\mathcal{F}}(\cdot)$ is the $\mathtt{Softplus}$ activation function. The test-condition information is element-wise added to the state prediction network \cite{He2016_resnet}
\begin{equation}
\Vec{Z}_{i}^N  = \sigma_{\mathcal{F}} \circ L \left( \Vec{Z}_i^0 + \Vec{X}_{\mathcal{O}_N} \right) 
\end{equation}
The state of capacity is obtained by
\begin{equation}
Q^N_i =  \sigma_{\mathcal{F}} \circ L \circ \sigma_{\mathcal{F}} \circ L (\Vec{Z}_{i}^N)
\end{equation}
where $Q^N_i$ is the capacity for the $N$-th cycle (including the first cycle). Because the discharge capacity is always positive, $\sigma_{\mathcal{F}}$ is added to constrain the predicted capacity to be positive and accelerate the training process. $dQ/dV$ for the redox potential can be obtained via \texttt{PyTorch} auto-differentiation \cite{paszke2019pytorch}
\begin{equation}
    \left.\frac{dQ}{dV}\right|_i = \text{AutoDiff}(Q_i, V_i).
\end{equation}

\subsection{Model training}
The model is trained to minimize the sum of multi-task losses for the capacity of the first cycle, the $n$-th cycle, and $dQ/dV$:
\begin{equation}
    \mathcal{L} =  w_Q\ell(Q_i^N) + w_{dQ}\ell(\frac{dQ^N}{dV_i}) + w_{Q_1}\ell(Q_i^1)  + \mathcal{R}.
    \label{eq:loss_function}
\end{equation}
A \texttt{MSE} loss function is used for $\ell(Q_i^N)$ and $\ell(\frac{dQ^N}{dV_i})$, whereas a \texttt{MAE} loss function is employed for the first cycle as a contrastive term $\ell(Q_i^1)$. The weights for $Q_i^N$, $dQ/dV$, and $Q_i^1$ are set to $w_Q$ = 1, $w_{dQ}$ = 1, and $w_{Q_1}$ = 5. The term $\mathcal{R}$ represents regularization, which consists of two parts: (1) an $\ell_2$-norm regularization of the network's parameters $||\boldsymbol{\theta}||_2$ and (2) a smoothing term $||dQ/d\textbf{c}||_2$  to avoid large, unphysical performance fluctuations ($\textbf{c}$ denotes the fractional concentration of elements). The weight of regularization is $10^{-4}$.

To make predictions, an ensemble of five independent models was trained to make predictions. Each model was trained with a batch size of 1024 within 30 epochs. The \texttt{Adam} optimizer was used with $10^{-3}$ as the initial learning rate. The \texttt{ExponentialLR} scheduler was used to adjust the learning rate with a decay of 0.9 per epoch.

\section{Acknowledgments}
This work was primarily supported by the U.S. Department of Energy, Office of Science, Office of Basic Energy Sciences, Materials Sciences and Engineering Division under Contract No. DE-AC0205CH11231 (Materials Project program KC23MP). The data collection in this work was supported by the Assistant Secretary for Energy Efficiency and Renewable Energy, Vehicle Technologies Office, under the Advanced Battery Materials Research (BMR) Program of the US Department of Energy (DOE) under contract No. DE-AC0205CH11231. The computational modeling in this work was supported by the computational resources provided by the Extreme Science and Engineering Discovery Environment (XSEDE), supported by National Science Foundation grant number ACI1053575; the National Energy Research Scientific Computing Center (NERSC); and the Lawrencium computational cluster resource provided by the IT Division at the Lawrence Berkeley National Laboratory. The authors thank Huiwen Ji, Jianping Huang, and Zijian Cai for their help in experimental data collection, and Yifan Chen for valuable discussions.

\section{Availability}
The codes of \texttt{DRXNet} are open-sourced at \href{https://github.com/zhongpc/DRXNet}{https://github.com/zhongpc/DRXNet} and \href{https://doi.org/10.5281/zenodo.10719829}{https://doi.org/10.5281/zenodo.10719829}. The open-source dataset is available at \href{https://doi.org/10.6084/m9.figshare.25328578.v1}{https://doi.org/10.6084/m9.figshare.25328578.v1} for public access, which contains 12,688 experimental discharge voltage profiles excluding the Mn-rich and Ti-based DRX. The open-source dataset is not identical to, but rather a part of, the DRX-TD that was used for the pretrained models in the paper.

\section{Supplementary Information}
\label{sec:SI}
Supplemental information can be found online at \href{https://doi.org/10.1016/j.joule.2024.03.010}{https://doi.org/10.1016/j.joule.2024.03.010}.

\section{Author Contributions}
P.Z. and G.C. conceived the initial idea. P.Z. collected the dataset and developed the code base with help from B.D. and T.H.. Z.L. and G.C. offered insight into the project. P.Z. and G.C. prepared the manuscript. All authors contributed to discussions and approved the manuscript.

\bibliography{references}
\end{document}